\documentstyle[12pt]{article}

\begin{document}
\hfill hep-th/yymmnn
\begin{center}
{\Large\bf The von Neumann-Wigner type potentials and the wave
functions' asymptotics for the discrete levels in continuum}

\medskip
\medskip
    {\bf A.\ Khelashvili and N.\ Kiknadze}

\medskip

{\it High Energy Physics Institute and Dept.\ of General Physics,
Tbilisi State University, Chavchavadze ave.\ 1, Tbilisi 380028, Georgia.}
\end{center}

\medskip
\begin{abstract}
One to one correspondence between the decay law of the
von Neumann-Wigner type potentials and the asymptotic behaviour of
the wave functions representing bound states in the continuum is
established.
\end{abstract}

Many years ago von Neumann and Wigner \cite{1} discovered a class
of potentials that gives isolated quantum mechanival levels embedded
in continuum of positive energy states. The underlying strategy of
these authors was employed in \cite{2} to produce more examples. The
main feature of these potentials is oscillations at the spatial infinity
together with a relatively slow decrease. A lot of authors have
contributed to the sound mathematical substantiation of this extraordinary
phenomenon and most of their results are collected in the excellent
books of M.\ Reed and B.\ Simon \cite{3}. Recently the interest in
this problem was excited anew due to its various applications in
physics of atoms and molecules. In spite of rather wide
publications, as it was noted in the recent review \cite{4}, ``there
is not as yet a fully systematic approach". In \cite{4} the
isospectral technique is applied to generate such kind of
potentials.

One of the most fundamental conclusions of all previous
investigations is that for so called modulating functions
constructed or used there, the normalizable (square integrable) wave
functions have only a power-like decay at large distances while
potentials vanish in the same limit. Hence, these wave functions can
hardly be called bound states in the usual sense, because in general
they do not guarantee finiteness even of the square radius of the
state.

Below we present a slightly modified, but by our opinion more
convenient method that allows to observe one to one correspondence
between the decay law of potentials and that of wave functions
corresponding to the bound states in continuum.
Moreover we will demonstrate that
there exist potentials, which lead to wave functions with more
rapid than the power-like decrease.

We confine ourselves to considering of S-waves only. Corresponding
Schr\"o\-dinger equation for radial function $\chi$ has the form
\begin{equation}
\chi''(r) + \frac{2m}{\hbar^2}\left[E-U(r)\right]\chi(r) = 0.
\label{1}
\end{equation}

Denoting
\begin{equation}
\frac{2mE}{\hbar^2} = k^2, \qquad \frac{2mU(r)}{\hbar^2} = V(r)
\label{2}
\end{equation}
we find from eq.(\ref{1}) that
\begin{equation}
V(r) = k^2 + \frac{\chi''}{\chi}.  \label{3}
\end{equation}
Now, following \cite{1}\cite{2} we take
\begin{equation}
\chi(r) = \chi_0(r) f(r), \label{4}
\end{equation}
where $\chi_0(r)$ is solution of some solvable Schr\"odinger
equation and $f(r)$ is a modulating function. As a rule the free or
Coulomb solutions are used for $\chi_0$ \cite{1}\cite{2}\cite{4}
and the boundary condition at the origin $\chi(0)=0$ is
satisfied by a suitable choice of them. As an example let us take the
free solution:
\begin{equation}
\chi_0(r)=\frac{1}{k}sin(kr). \label{5}
\end{equation}

After substituting (\ref{4})--(\ref{5}) into eq.(\ref{3}) one
obtains \cite{2}
\begin{equation}
V(r) = \frac{f''}{f} + 2k\frac{f'}{f}ctg(kr). \label{6}
\end{equation}

We must choose the function $f(r)$ in a manner to provide
cancellation  of poles of $ctg(kr)$, i.e.\ the zeroes of $sin(kr)$.
Usually it is achieved by taking $f(r)$ to be differentiable
function of the variable \cite{1}\cite{2}:
\begin{equation}
s(r)=k\int_{0}^{r} sin^2(kr')dr'=\frac{1}{2}kr-\frac{1}{4}sin(2kr).
\label{7}
\end{equation}

Instead of setting the function $f(r)$, we will set its logarithmic
derivative
\begin{equation}
{\cal C}(r)\equiv\frac{f'}{f}.  \label{8}
\end{equation}
Then
\begin{equation}
V(r)={\cal C}^2(r)+{\cal C}'(r)+2kctg(kr){\cal C}(r) \label{9}
\end{equation}
and the modulating function $f(r)$ can be constructed by solving
eq.(\ref{8}):
\begin{equation}
f(r)=A exp\left\{\int_0^r {\cal C}(z)dz\right\}.
\label{10}
\end{equation}

First of all we must take care for the above mentioned cancellation
of the poles. We can take
\begin{equation}
{\cal C}(r)=\phi(r)sin^2(kr). \label{11}
\end{equation}
Then the potential becomes
\begin{equation}
V(r)=\phi^2(r)sin^4(kr)+\phi'(r)sin^2(kr)+2k\phi(r)sin(2kr).
\label{12}
\end{equation}
Next, to obtain potential that vanishes at the spatial infinity we
probe
\begin{equation}
\phi(r)=\frac{a}{r^\beta}, \qquad a=const, \qquad \beta>0.
\label{13}
\end{equation}
So the potential takes the form
\begin{equation}
V(r)=\frac{a^2sin^4(kr)}{r^{2\beta}}-\frac{a\beta sin^2(kr)}{r^{1+\beta}}
+\frac{2aksin(2kr)}{r^{\beta}}
\label{14}
\end{equation}
and the corresponding modulating function is
\begin{equation}
f(r)=A exp\left\{a\int_0^r \frac{sin^2(kz)}{z^\beta}dz\right\}.
                                       \label{15}
\end{equation}

Evidently, if $\beta>0$ the last term will dominate in (\ref{14}) as
$r\to\infty$. According to the theorem $XIII.58$ from \cite{3} there
are no normalizable wave functions for positive eigenvalues if
$\beta>1$. Validity of this theorem in our case can be checked
immediately by studying asymptotic behaviour of (\ref{15}). Therefore
only $\beta\leq 1$ case is of interest. Let us take
$\beta=1-\epsilon$ with $\epsilon>0$ ($\epsilon=0$ case must be
considered separately). We have
\begin{equation}
\int_0^r \frac{sin^2(kz)}{z^\beta}dz=\frac{r^\epsilon}{2\epsilon}-
\frac{1}{4}\left[\frac{\gamma(\epsilon,2ir)}{(2i)^\epsilon}+
\frac{\gamma(\epsilon,-2ir)}{(-2i)^\epsilon}\right].
\label{16}
\end{equation}
Here $\gamma(a,x)$ denotes the incomplete gamma function \cite{5},
which has convergent series in positive powers of $x$
\begin{equation}
\gamma(a,x)=e^{-x}\sum_{n=0}^{\infty}\frac{x^{a+n}}{(a)_{n+1}}
\label{17}
\end{equation}
and the asymptotic expansion in inverse powers of $x$
\begin{equation}
\gamma(a,x)=\Gamma(a)+x^{a-1}e^{-x}\left[\sum_{m=0}^{M-1}
\frac{(1-a)_m}{(-x)^{m}}+O\left(\mid x\mid^{-M}\right)\right].
\label{18}
\end{equation}
It follows that in the limit $r\gg 1$
\begin{equation}
\gamma(\epsilon,2ir)\approx \Gamma(\epsilon)-(2ir)^{\epsilon-1}
e^{-2ir}+O\left(r^{\epsilon-2}\right).
\label{19}
\end{equation}
Therefore
\begin{equation}
\int_0^r
\frac{sin^2(kz)}{z^{1-\epsilon}}dz\approx\frac{r^\epsilon}{2\epsilon},
\qquad \qquad r\gg 1
\label{20}
\end{equation}
and if we take coefficient $a$ in (\ref{15}) to be negative, then
$f(r)$ would have quasi-exponentially decreasing asymptotics leading
to square integrable wave functions. Moreover, according to (\ref{17})
the modulating function $f(r)$ tends to constant as $r$ approaches
origin and does not destroy correct boundary behaviour of the wave
function $\chi$.

Collecting all above results together we conclude that if the
potential has the dominating asymptotics like
\begin{equation}
V(r)\sim -\frac{2\mid a\mid k}{r^\beta}sin(2kr),
\qquad\qquad 0<\beta<1
          \label{21}
\end{equation}
then the wave function $\chi$ behaves like
\begin{equation}
\chi(r)\sim sin(kr) exp\left\{-\frac{\mid a\mid
r^{1-\beta}}{2(1-\beta)}\right\}, \qquad \qquad r\gg 1
\label{22}
\end{equation}
and so decreases fast enough to be normalizable.

Let us consider now the limiting case $\beta=1$ and define
\begin{equation}
I=\lim_{\sigma\to 0}\int_\sigma^r\frac{sin^2(kz)}{z}dz=
\frac{1}{2}\left(ln(kr)-{\it Ci}(2kr)+\gamma+ln2\right),
\label{24}
\end{equation}
where $\gamma$ is the Euler constant and {\it Ci}$(u)$ --- the integral
cosine, which has the following asymptotics \cite{5}
\begin{eqnarray}
{\it Ci}(u)  \approx  \gamma+ln(u)-\frac{u^2}{4}+O\left(u^4\right)
& &  u\ll 1 \nonumber \\
{\it Ci}(u)  \approx  sin(u)+O\left(u^{-1}\right)
& & u\gg 1 \label{25}
\end{eqnarray}
Therefore
\begin{eqnarray}
I  \longrightarrow  \frac{1}{2}(kr)^2, & & kr\ll 1
\nonumber \\
I  \longrightarrow  \frac{1}{2}ln(kr), & &  kr\gg 1
\label{26}
\end{eqnarray}

Making use of (\ref{25})--(\ref{26}) in (\ref{24}) and then in
(\ref{15}), we see that
\begin{eqnarray}
f(r)  \longrightarrow  const, & & kr\ll 1
\nonumber \\
f(r)  \longrightarrow  \left(kr\right)^{{}^a/{}_2},
& &  kr\gg 1
\label{27}
\end{eqnarray}
and therefore
\begin{equation}
\chi(r)\longrightarrow r^{-{}^{\mid a\mid}/{}_{2}} sin(kr),
\qquad\qquad kr\gg 1
\label{28}
\end{equation}
It unifies correctly all known results derived for the
$r^{-1}sin(2kr)$ asymptotic behaviour of the potential and agrees
with the Atkinson's theorem \cite{3}.

As a conclusion we can say that there is one to one correspondence
between the asymptotic behaviour of potentials decreasing with
oscillations and that of wave functions belonging to bound
states in continuum. A slight modification of point of view (see
eqs.(\ref{8})--(\ref{12})) allowed us to yield generalized von
Neumann-Wigner type potentials with the arbitrary powers of
decrease, $\beta\neq 1$. Only $\beta\leq 1$ gives bound states in
continuum.
Of course the correspondence found above between the
asymptotics does not depend on
the method of construction --- it is general because its validity
depends only on the asymptotic behaviour of the potential under
consideration. The last comment we want to make is that the
pure exponential decrease $exp(-|a|r/2)$ of
the wave function corresponds to potentials
that do not vanish at the infinity, but only oscillate (the case
$\beta=0$). The
finiteness of any characteristic dimensions of the bound state
(like square radius or any higher moments of $r$) makes principal difference
between the found new solutions (for $\beta\neq 1$) and the
known ones
($\beta=1$), namely the former are localized like ordinary bound states,
while the latter are not.


\begin{thebibliography}{99}
\bibitem{1}
J. von Neumann and E. Wigner, Phys. Z. {\bf 30} (1929) 465.
\bibitem{2}
F. H. Stillinger and D. R. Herrik, Phys. Rev. {\bf A11} (1975) 446.
\bibitem{3}
M. Reed and B. Simon, {\it Methods of Modern Mathematical Physics},
volumes 3 and 4. Acad. Press, New-York-San Francisco, 1975.
\bibitem{4}
F. Cooper, A. Khare and U. Sukhatme, Phys. Reports, {\bf 251}, No 5
and 6 (1995).
\bibitem{5}
H. Bateman and A. Erdelyi, {\it Higher transcendental functions},
vol. 2, N.Y. (1953).
\end{thebibliography}
\end{document}